\documentclass[letterpaper,titlepage,11pt]{article}

\pdfoutput=1

\usepackage{amsmath}
\usepackage{amsfonts}
\usepackage{amssymb}
\usepackage{graphicx} 
\usepackage{subfigure}
\usepackage{enumerate}
\usepackage[
      colorlinks=true,
      linkcolor=blue,
      urlcolor=blue,
      filecolor=black,
      citecolor=red,
      pdfstartview=FitV,
      pdftitle={},
        pdfauthor={Roberto Emparan},
        pdfsubject={},
        pdfkeywords={},
        pdfpagemode=None,
        bookmarksopen=true
      ]{hyperref}

\def\clock{{\count0=\time
           \divide\count0 60
           \ifnum\count0<10 0\fi\the\count0
           \multiply\count0 -60 \advance\count0 \time
           :\ifnum\count0<10 0\fi \the\count0
         }}
\newcommand{\timestamp}{{\small\vbox{\hbox{\tt\jobname.tex}
\hbox{\the\day/\the\month/\the\year, \clock}}}}

\newcommand{\ie}{{i.e., }}
\newcommand{\eg}{{e.g., }}

\newcommand{\lp}{\left(}
\newcommand{\rp}{\right)}

\newcommand{\beq}{\begin{equation}}
\newcommand{\eeq}{\end{equation}}
\newcommand{\beqa}{\begin{eqnarray}}
\newcommand{\eeqa}{\end{eqnarray}}

\begin{document}
\begin{titlepage}
\leftline{}
\vskip .7cm
\centerline{\LARGE \bf Predictivity lost, predictivity regained:}
\bigskip
\centerline{\LARGE \bf a Miltonian cosmic censorship conjecture}
\vskip 1.1cm
\centerline{\bf Roberto Emparan}
\vskip 0.5cm\centerline{\sl  Instituci\'o Catalana de Recerca i Estudis
Avan\c cats (ICREA)}
\centerline{\sl Passeig Llu\'{\i}s Companys 23, E-08010 Barcelona, Spain}
\smallskip
\centerline{and}
\smallskip
\centerline{\sl Departament de F{\'\i}sica Qu\`antica i Astrof\'{\i}sica, Institut de
Ci\`encies del Cosmos,}
\centerline{\sl  Universitat de
Barcelona, Mart\'{\i} i Franqu\`es 1, E-08028 Barcelona, Spain}
\smallskip
\vskip 0.5cm
\centerline{\small\tt emparan@ub.edu}

\vskip .6cm
\centerline{\bf Abstract} \vskip 0.3cm 
\noindent 
Cosmic censorship is known to fail in some well-controlled phenomena, calling into question  the predictive power of General Relativity and opening up the possibility of observing Planck-scale physics. We propose that the cosmic censorship conjecture can be amended so that its spirit prevails. Naked singularities that, classically, have zero mass are allowed. Physically, these are Planck-sized `black holes', which evaporate in a few Planck times. General Relativity fails only for a tiny interval in time, to then quickly regain control in a Miltonian evolution that returns us to the predictive paradise of Einstein's equations. If this refinement of the conjecture is correct, then, even though Nature does allow to expose breakdowns in the smooth fabric of spacetime, it limits them to a mostly harmless minimum.

\vfill

\centerline{\small Essay written for the Gravity Research Foundation 2020 Awards for Essays on Gravitation}
\centerline{\small Submitted March 31, 2020}

\end{titlepage}
\pagestyle{empty}
\small
\normalsize
\newpage
\pagestyle{plain}
\setcounter{page}{1}

The cosmic censorship conjecture asks whether a system following the classical equations of Einstein can eventually lay bare a disruption in the smooth fabric of spacetime, giving way to a fully quantum gravitational description. It is also a question about the in-principle feasibility of the ultimate `gravitational collider', by letting the crushing force of gravity run its course towards the highest energy densities and curvatures -- and in a way that we can observe the new physics safely from afar. 

More than fifty years after it was proposed \cite{Penrose:1969pc}, we now know that cosmic censorship fails to banish the large, naked curvatures that appear in certain well-controlled situations that we discuss below. But, what do these examples teach us about the predictive power of General Relativity, and about the observability of Planck-scale physics?

In this essay we argue that, although the examples do show that cosmic censorship is disobeyed in letter, they also indicate that, with a natural amendment, it prevails in spirit. Predictivity of General Relativity can be lost, but only for a tiny interval in time and involving tiny amounts of energy, with the classical theory quickly regaining control:  a Miltonian evolution that restores the predictive paradise of Einstein's theory. 

Let us unwrap more carefully these statements, keeping the physics perspective in mind. We are referring to so-called weak cosmic censorship, the conjecture that smooth initial data with physically reasonable matter evolve under Einstein's  equations without developing naked singularities. This is concerned with what can be seen from a long distance \ie at null infinity. We will not dwell on strong cosmic censorship, whose main concern is the fate of the black hole interior, with different physical implications than we discuss here.

In mathematical formulations, the term `singularity' is replaced with the notion of incompleteness of the evolution. While this captures the idea of lack of predictivity, it leaves us in the dark about why the evolution stops.
We will assume, with Penrose, that ``the most “reasonable” explanation (\dots) seems to be that the space-time is confronted with, in some sense, infinite curvature at its boundary.''\cite{Penrose:1969pc} It is also the most interesting possibility: if the naked curvature grows arbitrarily large, then it will necessarily reach Planck-scale values exposing quantum gravity to observation. For us, the singularity will not be a place of `infinite curvature', but a synonym for a Planckian region in an otherwise smooth spacetime.

\paragraph*{Strength of violations.} Say that an initially smooth system evolves to form a singularity that concentrates a fraction $M_\text{sing}/M$ of the initial mass. Without a general understanding of singularities, it is difficult to give a precise and general enough definition of their mass $M_\text{sing}$, but we can consider a spatial slice along the evolution, where the curvature is growing unbounded within a region that we surround with a surface where a suitable quasilocal notion of mass can be computed, e.g., the Brown-York energy or the Hawking mass. Although this might not be applied to all conceivable singularities, it will do for the naked ones that we conjecture are allowed. If $M_\text{sing}$ cannot be defined in any sensible way, then it likely means that our conjecture does not hold.

We now distinguish between strong violations, with $M_\text{sing}\sim M$, \ie most of the initial mass goes into Planckian densities, and mild violations, with $M_\text{sing}\sim M_\textrm{Planck}$,\footnote{Possibly also $M_\textrm{sing}\sim (M_\textrm{Planck})^\alpha$ with $0<\alpha\leq 1$.} where only a tiny mass gets into the quantum gravity regime -- somewhat analogous to the LHC, where trillions of collisions are needed to form a single new particle. Since in the classical limit $M_\textrm{Planck}\to 0$, mild violations result when the mass of the classical singularity vanishes.

\paragraph*{Time scales.} We have known for long how to reach Planckian curvatures starting from arbitrarily low ones: let a large black hole evaporate through Hawking emission until its mass is $M\sim M_\textrm{Planck}$. The time this takes
\beq
t_\textrm{evap}\sim t_\textrm{Planck} \lp \frac{M}{M_\textrm{Planck}}\rp^3
\eeq
is a long quantum time scale that diverges as $\hbar\to 0$, that is, as the Planck-scale cutoff is removed. Cosmic censorship asks that the evolution is instead classically-driven, therefore much faster, within the typical classical time
\beq\label{tclass}
t_\textrm{class} \sim \frac{GM}{c^3}.
\eeq
For a solar mass, $t_\textrm{class}\sim 10^{-5}$ seconds, while $t_\textrm{evap}\sim 10^{65}$ years: a classically-driven violation is more time-efficient for exploring Planckian physics than quantum evaporation. Whether we call the latter a violation of cosmic censorship or not (and we usually do not) is semantics.
The lesson is to distinguish between singularities formed in a time that remains finite as $M_\textrm{Planck}\to 0$, and those for which it diverges. In the remainder we will only discuss classically-driven fast violations.

\paragraph*{Cosmic censorship violations.} There are two well attested paradigms for violations of cosmic censorship: Choptuik's critical collapse \cite{Choptuik:1992jv}, and black string pinches in the evolution of the Gregory-Laflamme instability \cite{Gregory:1993vy,Lehner:2010pn}.\footnote{Other proposed mild violations \cite{Eperon:2019viw} also fit in our framework.}

\begin{figure}[t]
 \begin{center}
  \includegraphics[width=\textwidth,angle=0]{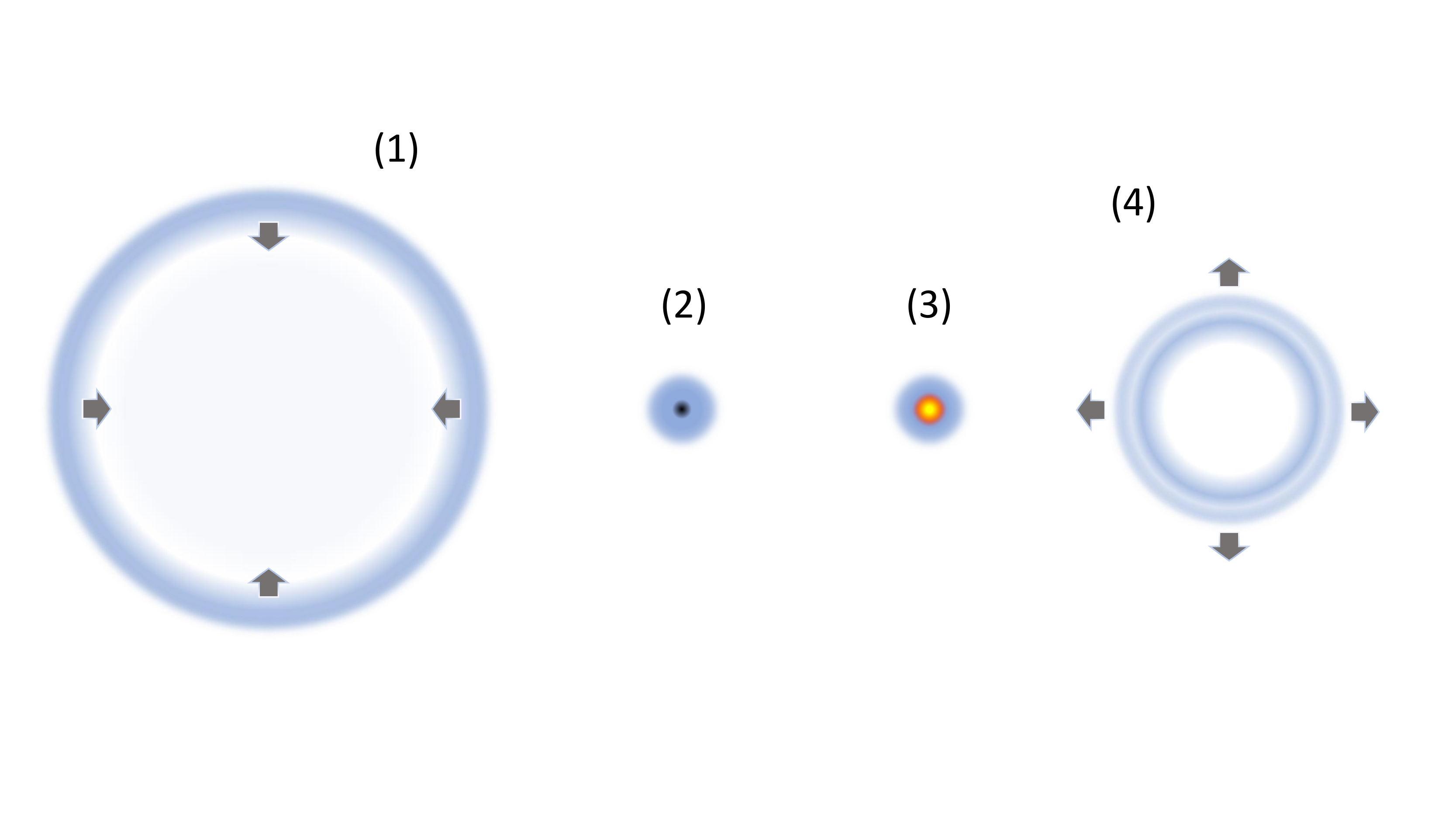}
 \end{center}
 \vspace{-20mm}
 \caption{\small Proposed evolution of critical collapse. At (3), the tiny Planck-sized `black hole' quickly decays via quantum effects. The rest of the evolution is classical.}
 \label{fig:Collapse}
\end{figure}

Consider a spherical scalar field cloud collapsing under the action of gravity  (fig.~\ref{fig:Collapse}). If its initial amplitude $a$ is small, it contracts and bounces back away to infinity. For large amplitudes, it forms a black hole. We can tune $a$ to a regime where the mass of the resulting black hole scales as \cite{Choptuik:1992jv}
\beq
M\propto (a-a_*)^{0.374}\,.
\eeq
As $a\to a_*$, the solution approaches a naked singularity with zero mass. Physically, it forms a Planck-sized `black hole' with mass $\sim M_\textrm{Planck}$: a mild violation.

\begin{figure}[t]
 \begin{center}
  \includegraphics[width=\textwidth,angle=0]{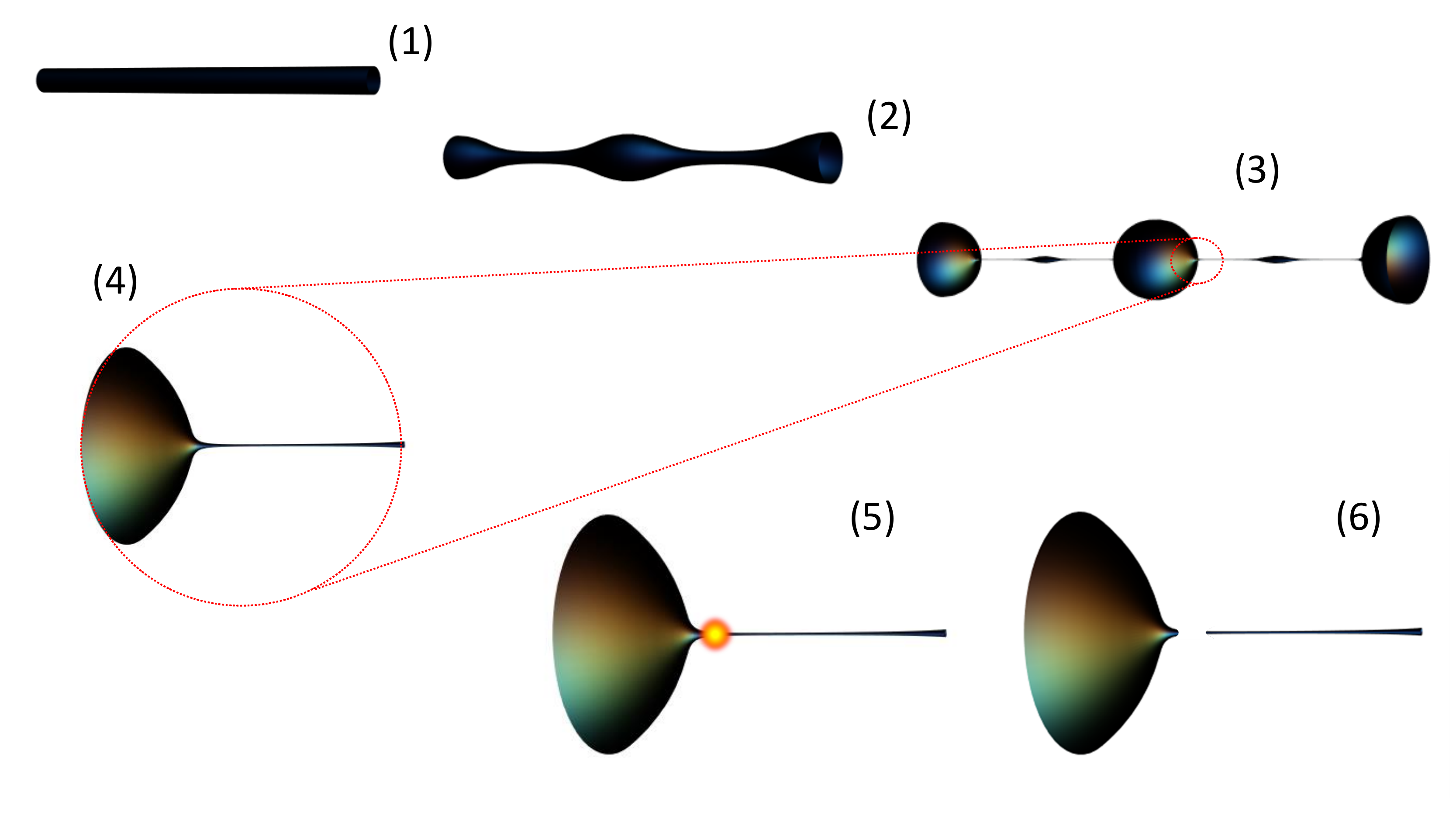}
 \end{center}
 \vspace{-8mm}
 \caption{\small Proposed evolution of black string pinches: only at (5) does quantum gravity enter to `evaporate' the Planckian neck. The rest is governed by the classical Einstein equations. Fluid jets break up this way \cite{Eggers:1997zz}, with General Relativity $\leftrightarrow$ Hydrodynamics, and Quantum Gravity $\leftrightarrow$ Molecular dynamics.}
 \label{fig:BSEvol}
\end{figure}

Black strings are solutions of the Einstein equations constructed by extending the Schwarzschild geometry along an extra dimension. 
Ref.~\cite{Lehner:2010pn} evolved numerically a perturbed black string and proved that it behaves much like a fluid jet: it grows blobs joined by thinning tubes around which the curvature grows unbounded. The tubes reach Planck size in necks, in a mild violation of cosmic censorship. Similar singular pinches appear in higher-dimensional black hole collisions \cite{Andrade:2018yqu,Andrade:2019edf}.  No fine tuning is needed.

\paragraph*{Predictivity lost, predictivity regained.}

Some purported violations of cosmic censorship involve unphysical matter, \eg pressureless dust. Not these ones: scalar fields and gravity are definitely physical.
Black strings need additional dimensions, but these are within experimental bounds and well motivated by string theory.
Critical collapse is often dismissed as an unphysical violation due to the fine-tuning required, but the actual distinction is that of experimental physics vs.\ observational physics. Although very hard to observe in the skies by astronomers, collapse to a Planck-sized `black hole' may well be engineered by the experimentalists of an advanced civilization (and the fine-tuning is a moderate power law in the initial data).

So, rather than modifying \emph{ad hoc} the rules of the game to ban these examples, we will accept that cosmic censorship is violated and learn what this implies.
Must we wait for a fully fledged quantum theory of gravity to tell us how to continue the evolution? We do not think so: a plausible resolution of mild singularities is available with minimal input from quantum gravity.

Consider first the neck in the black string pinch. It may be regarded as a Planck-size `black hole' with very high effective temperature, which, without any conserved charges that could prevent it, will quantum-mechanically decay by emitting a few Planckian quanta in  a time $\sim t_\textrm{Planck}$. That is, it evaporates like the neck in a fluid jet does (literally) and breaks the jet into droplets. And just like, afterwards, hydrodynamics quickly resumes control of droplet evolution, General Relativity will describe the resulting horizons moving apart. Note that in the classically-driven evolution there is no black hole unitarity paradox.

Similarly, in critical collapse the Planckian mass `black hole'  will evaporate in a time $\sim t_\textrm{Planck}$, emitting only a few quanta of energy $\sim E_\textrm{Planck}$. The large scalar cloud will dissipate away and the geometry will relax  following the classical field equations, almost unaffected by the tiny evaporation episode.

In both cases the loss of classical predictivity is very small: the uncertainty in the evolution after the evaporation will be proportional to at most a power of  $(M_\textrm{Planck}/M)$. Predictivity of the entire evolution using General Relativity will be restored to great accuracy.

We are then led to formulating an amendment to the cosmic censorship conjecture: 
\begin{itemize}
\item Mild violations, and only them, are allowed; classically, the singularities have zero mass.
\end{itemize}
Furthermore, we propose that mild singularities are resolved with a little `quantum gravity pixie dust' that makes the Planck-sized object evaporate in a time $\sim t_\textrm{Planck}$. So, classically, this is a zero mass object lasting zero time. An interesting challenge is to fix in a mathematically precise and unique way the evolution of Einstein's equations across these mild singularities -- a problem possibly similar to evolving the Navier-Stokes equations across shocks -- to fully return the predictivity of General Relativity. 

\paragraph*{Charged violations and weak gravity.}

What if the Planck-sized `black hole' had a conserved charge $Q=M_\textrm{sing}$? This could stabilize it and prevent its evaporation, much like extremally charged black holes do not emit Hawking radiation. Such a long-lived singularity -- a Planckian remnant, but without information problems -- could play havoc with the classical evolution. The danger is dispelled if the `weak gravity conjecture' \cite{ArkaniHamed:2006dz} holds, that is, if there necessarily exist particles with $q>m$ which provide a swift decay channel for the `extremal singularity'. Interestingly, this conjecture has already been invoked to ban a potentially strong (classically-driven but in a time diverging as $M_\textrm{Planck}\to 0$) charged violation of cosmic censorship \cite{Horowitz:2016ezu,Crisford:2017gsb}. 

\bigskip

So, can classical gravity bring about observable breakdowns in the geometry of spacetime? It seems so, but -- if our conjecture is correct -- only in a minimal way. Nature appears to much more favor the rule of Einstein's equations, only slightly punctuated.

\medskip

\paragraph*{Acknowledgments.} These ideas (partly published in \cite{Andrade:2018yqu,Andrade:2019edf}) have been presented over the years at many meetings (``Holography, black holes and numerical relativity'' Southampton (2017); ``Quantum Physics and Gravity'' Vienna (2017); EREP M{\'a}laga (2017); ``Quantum Black Holes in the Sky?'' Perimeter Institute (2017);  ``Big Questions in Physics'' Arizona (2019);  ``Black Hole Initiative Colloquium'' Harvard (2019); ``Weak Gravity and Cosmic Censorship'' Princeton (2019); ``Gravity workshop'' Benasque  (2019); ``Gravitational Holography'' KITP, UCSB (2020); and seminars at U.\ of Michigan, MIT, King’s College, and Imperial College London (2019)) and I would like to thank the audiences at all these places for their positive response and feedback.

Work supported by ERC Advanced Grant GravBHs-692951, MEC grant FPA2016-76005-C2-2-P, and AGAUR grant 2017-SGR 754.


\end{document}